# Joint Action is a Framework for Understanding Partnerships Between Humans and Upper Limb Prostheses

**Michael R. Dawson[1,2*], Adam S. R. Parker[2,3], Heather E. Williams[2,4], Ahmed W. Shehata[1], Jacqueline S. Hebert[1,4], Craig S. Chapman[5], Patrick M. Pilarski[1,2]**

[1]Division of Physical Medicine and Rehabilitation, Department of Medicine, University of Alberta, Edmonton, Alberta, Canada

[2]Alberta Machine Intelligence Institute (Amii), Edmonton, Alberta, Canada

[3]Faculty of Rehabilitation Medicine, Edmonton, Alberta, Canada

[4]Department of Biomedical Engineering, University of Alberta, Edmonton, Alberta, Canada

[5]Faculty of Kinesiology, Sport, and Recreation, University of Alberta, Edmonton, Alberta, Canada

**\* Correspondence:**
Michael R. Dawson
mrd1@ualberta.ca

## Abstract

Recent advances in upper limb prostheses have led to significant improvements in the number of movements provided by the robotic limb. However, the method for controlling multiple degrees of freedom via user-generated signals remains challenging. To address this issue, various machine learning controllers have been developed to better predict movement intent. As these controllers become more intelligent and take on more autonomy in the system, the traditional approach of representing the human-machine interface as a human controlling a tool becomes limiting. One possible approach to improve the understanding of these interfaces is to model them as collaborative, multi-agent systems through the lens of joint action. The field of joint action has been commonly applied to two human partners who are trying to work jointly together to achieve a task, such as singing or moving a table together, by effecting coordinated change in their shared environment. Using a joint action framework provides opportunities to understand the interactions between partners: how each represents the other's goal, their monitoring and prediction of each other's actions, the communication between them, and their ability to adapt to each other. In this work, we compare different prosthesis controllers—proportional electromyography with sequential switching, pattern recognition, and adaptive switching—in terms of how they present the hallmarks of joint action. The results of the comparison lead to a new perspective for understanding how existing myoelectric systems relate to each other, along with recommendations for how to improve these systems by increasing the collaborative communication between each partner.

## 1 Introduction

The development of robotic upper limb prostheses first began in the mid-twentieth century starting with single degree-of-freedom powered hands that were driven using voluntary surface electromyography (EMG) signals generated by residual muscles of the amputated limb (Parker et al.,

2006; Castellini et al., 2014). In these early systems, the prosthesis controller was calibrated in the clinic, fixed for day-to-day use, and considered as a standard example of human tool use. Since then, the capability and complexity of the prostheses has increased dramatically with additional powered joints and multi-articulated hands becoming available. With this scaling in both sensing and actuation technology, users continue to report challenges and frustration when controlling such a high number of joints with a small number of control inputs (Peerdeman et al., 2011; Resnik et al., 2014). To address user concerns, machine learning methods, such as pattern recognition (Micera et al., 2010; Scheme and Englehart, 2011; Shehata et al., 2021), deep neural networks (Williams et al., 2022), and continual learning methods like adaptive switching (Pilarski et al., 2012, 2013; Edwards et al., 2016a), have been developed to learn and specialize a device to the control signal patterns and daily life needs of individual prosthesis users (Castellini et al., 2014). These machine learning methods can be retrained or adapted during day-to-day use and generally delegate more autonomy to the prosthesis control system than earlier control approaches (Castellini et al., 2014). With the addition of more advanced computing technology, the traditional approach of modeling human-prosthesis interaction as a single human controlling a fixed tool has been suggested to no longer adequately capture the complex behavior of the co-adaptation between the human and the prosthesis controller (Schofield et al., 2021). We propose that our understanding of these more complex systems will improve by considering the human and the prosthetic device on comparable footing as partners working together to accomplish complex tasks, such as coordinated movement and object manipulation, through the lens of joint action.

*Joint action* can be defined as "any form of social interaction whereby two or more individuals coordinate their actions in space and time to bring about a change in the environment" (Sebanz et al., 2006). In the social sciences, joint action has been historically applied to two or more humans interacting together (Azaad et al., 2021), but more recently, studies have begun to investigate interaction between human and robots (Bicho et al., 2011; Grynszpan et al., 2019; Kathleen et al., 2022). Mathewson et al. (2022) introduced a framework for representing the prosthesis as a distinct agent (communicative capital). They used case studies and examples to outline the relationship between the agency of each partner in a human-prosthesis system and the capacity of the partnership. In the current article, we take these trends one step further by applying the viewpoint of joint action to human-prosthesis interaction (HPI). To accomplish this goal we explore how proportional EMG with sequential switching, pattern recognition, and adaptive switching controllers (a representative example of continual machine learning) fit into the framework, along with how these controllers might be further improved by incorporating additional hallmarks of joint action.

## 2 Joint Action

Several architectures of collaborative social interaction and joint action have been developed in the literature (Wolpert et al., 2003; Sebanz et al., 2006; Sebanz and Knoblich, 2009; Vesper et al., 2010; Knoblich et al., 2011; Pesquita et al., 2018). For the purposes of this paper, we build upon the architecture of Vesper et al. (2010) as it includes a clearly defined basic framework for evaluating whether a given system may develop into joint action. As per Vesper et al. (2010), there are four main hallmarks that should be present for a system to be considered as jointly acting: representations, monitoring, prediction, and coordination smoothers. Joint action also requires partners, tasks, and goals. We now define these terms and relate them specifically to the domain of HPI.

**Partners** – There needs to be at least two partners collaborating on a task for joint action to occur (Vesper et al., 2010). In the case of HPI, one partner is the *human partner* and the other the *machine partner* (in this case, we define the machine as the prosthesis and its control system).

**Tasks & Goals** - Each partner has an individual task that they are responsible for, which in combination, allows them to work towards accomplishing a shared goal (Vesper et al., 2010). The shared goal between the human and machine partner is to move the prosthesis to a particular position or use it to interact with the environment to bring about a specific, user-defined configuration of the environment. Typical tasks of the human include providing control signals to the machine partner and gross positioning of the prosthesis using body and residual limb movements (Major et al., 2014; Hebert et al., 2019). Typical tasks of the machine partner include interpreting the control signals from the human and selecting which motor(s) should be driven on the prosthesis.

**Representation** - As per Vesper et al. (2010), every partner working together should, at a minimum, internally depict for themselves their individual task and the shared goal (e.g., portrayals in the human brain or in digital storage). It may be helpful for each partner to have a representation of each other's individual task, but do not list this as a requirement.

**Monitoring** – In joint action, numerous perceptions or states are continually sampled or estimated by each partner, including their own actions and their partner's actions, along with how well they are achieving their individual tasks and shared goals (Vesper et al., 2010; Pesquita et al., 2018). The joint action architecture of Vesper et al. (2010) does not explicitly define which of these monitoring processes should be required, but instead indicates that necessary processes may be task-specific. In the context of HPI, we note that each partner will need to at least monitor their own actions, their partner's actions, and how well they are completing their own task. The human typically monitors the movement of the prosthesis via visual feedback. Conversely, the machine partner has an array of sensors, which may capture EMG signals from the human, as well as movements and forces from the prosthesis.

**Prediction** – Across multiple joint action architectures, each partner is considered to make predictions about their own actions or the actions of their partner (Vesper et al., 2010; Pesquita et al., 2018). These predictions could take many forms, including predicting the actions themselves, the timing of the actions, or the outcomes of the actions (Vesper et al., 2010).

**Coordination Smoothers** – Coordination smoothers are defined broadly as anything (besides the previously mentioned hallmarks) that improves coordination between the partners (Vesper et al., 2010). Examples of coordination smoothers that are particularly relevant to HPI include behaviors such as emphasizing actions and sending coordination signals. A complementary model of joint action (Pesquita et al., 2018) specifically mentioned the continuous improvement of predictions as a type of coordination smoother, which we have also included in our evaluation, due to its pertinence to the HPI setting.

## 3    Myoelectric Controllers Included in this Analysis

The three prosthesis controllers selected for analysis under the joint action architecture are each a representative case from three main categories of control: proportional EMG control with sequential switching (no machine learning), pattern recognition (batch/offline machine learning), and adaptive switching (continual/online machine learning) (Figure 1).

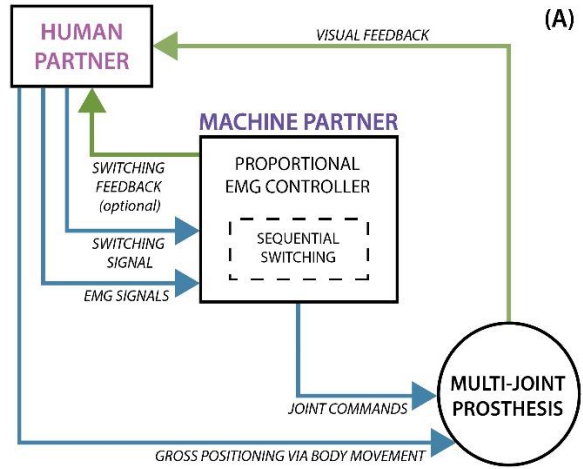
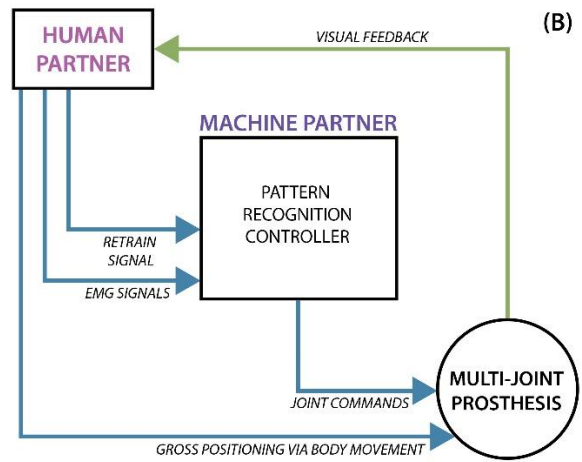
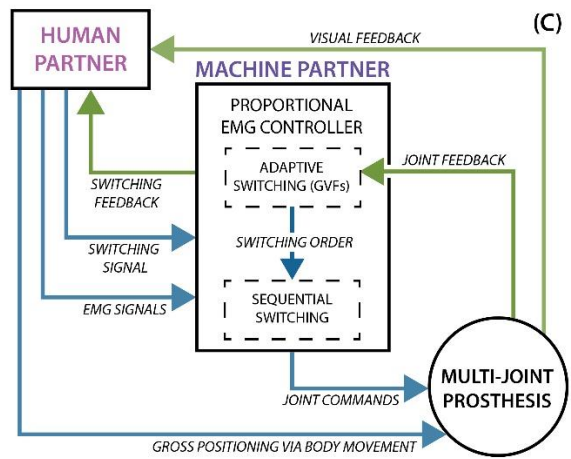

**Figure 1** – Block Diagrams of three different prosthesis controllers and how they interact with the prosthesis user. **(A)** Conventional myoelectric control with proportional EMG and sequential switching. **(B)** Pattern recognition controller with single class output and episodic (batched offline) training. **(C)** Adaptive switching, a continual learning method that dynamically adjusts the order of the sequential switching list. One common feature that is present in all three controllers is a physical connection between the prosthesis user and the multi-joint prosthesis via a prosthetic socket. EMG electrodes are integrated inside the socket, over residual muscles, and provide EMG signals to each prosthesis controller. The prosthesis user is also able to grossly position the prosthesis in space by moving their feet, trunk, or joints of their residual limb. Actuation of the motorized joints on the prosthesis is handled by the prosthesis controller. Another feature common to all three controllers is that the current state (e.g. position and velocity) of the multi-joint prosthesis is monitored by the human by visually attending to it.

### 3.1 Proportional EMG

The specific type of proportional EMG controller we focused on in this analysis is a two state proportional controller with sequential switching (hereafter shortened to proportional EMG) (Hubbard et al., 2004; Fougner et al., 2012), as illustrated in Figure 1 (A). EMG signals are acquired from residual antagonistic muscles (e.g. biceps/triceps), rectified and averaged, and then directly mapped to the velocity of a joint (e.g. hand open/close) on the prosthesis. After exceeding a threshold, the proportionality allows for slower or faster motor speeds, depending on the strength of the EMG signal, and the sequential switching allows the prosthesis user to sequentially switch between different joints on the arm (e.g. hand open/close, wrist rotation, elbow flexion). The switching signal is communicated from the human to the machine partner by co-contracting the two antagonistic muscles, pressing a button, or pulling on a cable switch.

### 3.2 Pattern Recognition

The pattern recognition controller (Scheme and Englehart, 2011), as illustrated in Figure 1 (B), is an *offline or batch machine learning* classifier that predicts the movement intent of the prosthesis user based on the pattern of their EMG signals. In contrast to proportional EMG controllers, pattern recognition controllers typically use additional EMG electrodes placed over muscles on the residual limb (Micera et al., 2010; Scheme and Englehart, 2011; Castellini et al., 2014; Shehata et al., 2021). Instead of relying solely on signal strength, Pattern recognition controllers extract multiple features from the time and frequency domains. Similar to proportional EMG control, pattern recognition controllers in contemporary clinical use are typically limited to controlling a single joint on the prosthesis at a time. The prosthesis user can initially train or retrain the classifier by momentarily activating the *retrain signal* (Fig. 1B*)*, which can be a button on the prosthesis or accessed via a phone application. A physical or virtual representation of the prosthesis will then move each joint one at a time in a prescribed manner while the human demonstrates their EMG signal pattern for each movement. After the demonstration period, the classifier will use the labeled samples to compute the parameters of the classifier that allow it to make predictions. This training process can take around a minute or two depending on the classifier. After training, the pattern recognition controller can be used in real-time by the prosthesis user to control the powered joints on their prosthesis.

### 3.3 Adaptive Switching

As an example of *continual or online machine learning* selected for this analysis, adaptive switching, as illustrated in Figure 1 (C), has a similar structure to a proportional EMG controller. However, instead of using a fixed switching list, adaptive switching uses the magnitude of a collection of learned general value functions (GVF, Sutton et al., 2011) to dynamically adapt the order of the switching list (Pilarski et al., 2012; Edwards et al., 2016a). GVF learning is a prediction approach, based on reinforcement learning methods, that can learn expected temporally extended accumulations of signals of interest based on a continuing stream of observations (Sutton et al., 2011; Pilarski et al., 2013), which in the adaptive switching case includes learned forecasts of prosthesis mode or function use (c.f., Pilarski et al., 2012). One key difference, compared to the other controllers, is that adaptive switching monitors the joint feedback from the prosthesis to make its predictions about which joint the human may want to use next. The improvement to the adaptive switching predictions happens continuously in real-time, during regular use, without the need for an explicit training period. Once a switching event is initiated by the human via a switching signal, the adaptive switching controller momentarily freezes the switching list while the human selects the joint and then resumes re-ordering once they start moving the joint (indicating that the correct joint was selected) (Edwards et al., 2016a).

## 4 Analysis and Discussion

The results of our analysis of the prosthesis controllers under a joint action lens are shown in Table 1. For all control schemes, the human partner was considered to reasonably demonstrate all of the hallmarks of joint action related to representations, monitoring, and predictions—humans demonstrating these hallmarks of joint action are supported by studies showing that prosthesis users adapt their internal models to take into account features of their prosthesis and its controller (Lum et al., 2014; Strbac et al., 2017; Blustein et al., 2018; Marasco et al., 2018). With respect to the machine partner, our analysis suggests that all three prosthesis controllers do exhibit many of the hallmarks of joint action, but that in all cases key hallmarks were missing or inadequate to fulfill the complete definition of joint action from Vesper et al. (2010). Joint action between a proportional EMG controller and the human is most likely not occurring since the controller lacks the necessary representation and prediction processes. In this case, it is more analogous to the common conception of prosthesis use as standard human tool use. If the shared goal can be internally captured by the pattern recognition and adaptive switching controller's representations, then it is likely that joint action may in fact be occurring in some way with the human partner during their use. However, if the shared goal is not within the scope of their representations, then joint action is arguably not occurring and we can consider it to be not expressly facilitated by their component machine learning processes independent of the capacity of their representations.

### 4.1 Representations, Monitoring, Tasks, and Goals

Monitoring, a consistent element of control engineering and human motor control, was prominent in regard to signals and states well within the observable space for a given partner (e.g., the machine partner monitoring the operation of its own inputs, outputs and state), but naturally less prominent for things that required more complete or nuanced representations of the full environment (e.g., partner and goals). Proportional EMG controllers employ fixed, direct mappings between inputs and outputs; such controllers do not have an explicit representation of any of the tasks or goals. However, it is less clear whether or not pattern recognition and adaptive switching have representations of their own tasks, the shared goal, or their partner's task. For pattern recognition, if the shared goal is to move to a particular position based on a pattern of EMG signals, then this controller may maintain some of these representations. However, if the shared goal is more complex and involves interacting with the environment (e.g. picking up an object), then from Figure 1 (B), we can see that the classifier does not have access to this type of information and likely would not have the required representations. The GVFs in adaptive switching do have access to position sensors and a load sensor in the gripper, which may help them infer the location of objects in the environment, so for these kinds of shared goals, the controller may have some of the required representations. However, if we abstract the shared goal to be a higher level task (e.g., folding a towel), then the controller likely does not have the required representations. We note that while not a candidate for our analysis in the present work, reinforcement learning algorithms with an externally provided reward signal would be well thought of as machine partners that represent and make decisions with respect to a goal (Pilarski et al., 2017; Mathewson et al., 2022).

**Table 1** – Evaluation of myoelectric controllers through a joint action lens. A green checkmark, yellow question mark, or red X symbol indicates that the controller presents, possibly presents, or does not present the listed hallmark.

| Hallmarks of Joint Action | Proportional EMG | | Pattern Recognition | | Adaptive Switching | |
|---|---|---|---|---|---|---|
| | Human Partner | Machine Partner | Human Partner | Machine Partner | Human Partner | Machine Partner |
| **Representation of:** | | | | | | |
| Own Task* | ✓ | X | ✓ | ? | ✓ | ? |
| Shared Goal* | ✓ | X | ✓ | ? | ✓ | ? |
| Partner Task | ✓ | X | ✓ | ? | ✓ | ? |
| **Monitor:** | | | | | | |
| Own Actions* | ✓ | ✓ | ✓ | ✓ | ✓ | ✓ |
| Own Task* | ✓ | ✓ | ✓ | ✓ | ✓ | ✓ |
| Partner Actions* | ✓ | ✓ | ✓ | ✓ | ✓ | ✓ |
| Partner Task | ✓ | X | ✓ | ? | ✓ | ? |
| Shared Goal | ✓ | X | ✓ | ? | ✓ | ? |
| **Predict:** | | | | | | |
| Own Actions | ✓ | X | ✓ | X | ✓ | X |
| Partner Actions* | ✓ | X | ✓ | ✓ | ✓ | ✓ |
| **Coordination Smoothers:** | | | | | | |
| Make Actions More Predictable | ✓ | X | ✓ | ✓ | ✓ | ✓ |
| Coordination Signals | X | ✓ | ✓ | ✓ | X | ✓ |
| Continuously Adapt predictions | ✓ | X | ✓ | X | ✓ | ✓ |

## 4.2 Predictions and Coordination Smoothers

We could not find evidence that any of the controllers were predicting aspects of their own actions. However, both pattern recognition and adaptive switching predict the movement intent of the human partner, which would help them better achieve the shared goal. With regard to coordination smoothers, we found evidence that humans make their actions more predictable by trying to generate more distinct EMG signals for all types of controllers, which can help improve their performance (Powell et al., 2014). Additionally, pattern recognition controllers often try to make their actions more predictable by mitigating the effects of incorrect predictions via techniques such as majority voting and velocity ramps (Englehart and Hudgins, 2003; Simon et al., 2011). Although adaptive switching is inherently less predictable than a stable ordered list, at critical moments it does make its actions more predictable by freezing the list when the human partner has triggered a switching event. Continuously adapting predictions are likely performed in all cases by the human as part of their learning process. Since the parameters for proportional EMG are typically only modified in the clinic, and given that the retraining of the classifier in pattern recognition only occurs intermittently, they do not learn continuously. Adapting and improving predictions in real-time are built into the regular operation of the GVFs in adaptive switching and so they demonstrated this hallmark. For coordination signals, we found that proportional EMG and adaptive switching both meet the condition by sending switching feedback to the human. However, the human does not send any explicit coordinating signals back to the machine partner. The coordination signals for pattern recognition mostly occur during the training phase, where the human sends the retrain signal and the pattern recognition visually displays the movements to guide the collection of EMG data.

## 5 Recommendations for Improving Human-Prosthesis Interaction

In the previous section, we analyzed which hallmarks of joint action were present in our different candidate human-prosthesis partnerships; whether joint action is occurring in these controllers is perhaps less important than what our analysis reveals in terms of recommendations for improving HPI by incorporating ideas from the field of joint action.

One powerful way to improve the HPI is to improve the representations or internal models of the human and prosthesis controller and by providing additional coordination smoothers. This could be done by increasing or improving the signals that each partner monitors about themselves or the task. This suggestion is supported by related literature— Shehata et al. (2018c) found that feedback about a control system to a human partner showed promise in increasing the strength of the internal model formed in the human brain about a prosthetic device and its control system, as well as the tasks performed. Specifically, feedback from the partner relayed through auditory, visual, and cutaneous sensory cues enabled the development of a strong internal model and allowed the human to better adapt to changes in the task and the prosthetic device (Shehata et al., 2018b). While task specific feedback to the human is known to improve performance for a given task (Engels et al., 2019), feedback from a machine partner about the control states and signals to the human partner has been shown to improve the human understanding of how a prosthetic device operates and thus improved the overall system integration (Shehata et al., 2018a). Achieving the optimal balance between providing information-rich feedback about how the machine partner is controlling the prosthetic device and task-specific feedback is key to improving the overall system performance without increasing the human's mental effort and information processing costs. Further opportunities also remain for exploring how subtle and ostensive cues and signaling from the human to the machine partner (either emergent conventions or pre-established conventions) can even further smooth their ongoing interactions (Scott-Phillips, 2015; Kalinowska et al., 2023).

A second pathway to improving HPI is by enhancing the ability of the partners to make predictions about themselves and each other. This could be accomplished either by strengthening predictive capabilities, blending of control solutions, or the use of additional sensors and communication channels to facilitate more comprehensive representations that can be used in contextually nuanced prediction learning. One such predictive enhancement has been made to adaptive switching through a technique called autonomous switching, which not only predicts what movement the prosthesis user wants to switch to next, but also *when* they will want to switch movements (Edwards et al., 2016b). With this method, when the autonomous switching controller reaches a minimum level of confidence, it will automatically switch (but will also allow the prosthesis user to manually switch if the controller guesses incorrectly). During the prediction process, autonomous switching can be seen to closely resemble the behavior of pattern recognition controllers, although through different computational implementations. Furthermore, the predictions from pattern recognition could be improved by increasing the contextual awareness of the classifier, through the use of additional sensors that describe the task or the human partner (Castellini et al., 2014; Shehata et al., 2021). As a recent example, data from multiple sensors have been used to facilitate precise classification of user-intended prosthesis movements for upper limb device control across multiple limb positions (Williams et al., 2022).

In conclusion, we believe joint action is a useful framework for thinking about and improving myoelectric control, and more broadly, for considering the human and their prosthesis as a dyad of interacting agents. This article contributed a focused analysis of the hallmarks of joint action, as presented by three representative examples of conventional myoelectric control, batch machine learning, and continual machine learning methods. It further presented two paths to improving myoelectric control through enhancing the capacity for coordination smoothing and strengthening the foundations for both human and machine partners to form and leverage predictive knowledge during their ongoing interactions. Based on this analysis, we recommend further study into the impact of joint action ideas on the wider field of neuroprosthetic control, and suggest that doing so will have a significant impact on the design and use of the next generation of individual-focused assistive rehabilitation technologies.

## 6  Author Contributions

PMP, and MRD contributed to the conception of the manuscript. MRD prepared the figure. All authors contributed to manuscript revision, read, and approved the submitted version.

## 7  Conflict of Interest

The authors declare that the research was conducted in the absence of any commercial or financial relationships that could be construed as a potential conflict of interest. We note that PMP is an employee of DeepMind, and that CSC is co-founder of the company Gaze and Movement Analysis Inc.

## 8  Acknowledgments

Research was supported by the Natural Sciences and Engineering Research Council of Canada (NSERC), Alberta Innovates, the Sensory Motor Adaptive Rehabilitation Technology (SMART) Network, the Alberta Machine Intelligence Institute (Amii), and the Canada CIFAR AI Chairs program.